\documentclass[preprint,showpacs,preprintnumbers,amsmath,amssymb]{revtex4}

\usepackage{graphicx}
\usepackage{graphics}
\begin{document}

\title{Queueing theoretical analysis of foreign currency exchange rates}
\author{Jun-ichi Inoue$^{1}$}
\email[e-mail: ]{j_inoue@complex.eng.hokudai.ac.jp}
\author{Naoya Sazuka$^{2}$}
\email[e-mail: ]{Naoya.Sazuka@jp.sony.com}
\affiliation{
$^1$Complex Systems Engineering, Graduate School of Information
Science and Technology, Hokkaido University, 
N14-W9, Kita-ku, Sapporo 060-0814, Japan \\
$^2$Sony Corporation, 
6-7-35 Kitashinagawa, Shinagawa, Tokyo 141-001, Japan
}
\begin{abstract}
We propose a useful approach for investigating the 
statistical properties of foreign currency 
exchange rates. 
Our approach is based on queueing 
theory, particularly, 
the so-called renewal-reward theorem. 
For the first passage processes of 
the Sony Bank 
US dollar/Japanese yen (USD/JPY) exchange rate, we 
evaluate the average waiting time 
which is defined 
as the average time that 
customers have to wait 
between any instant 
when they want to observe the rate 
(e.g. when they log in to their computer systems) and the next rate change. 
We find that the assumption of 
exponential distribution for the first-passage process 
should be rejected and that a Weibull distribution seems 
more suitable for explaining the stochastic process of the 
Sony Bank rate. 
Our approach also enables us 
to evaluate the expected reward for customers, 
i.e. one can predict how long 
customers must wait and how much reward 
they will obtain by the next price change 
after they log in to their computer systems. 
We check the validity of our 
prediction by comparing it with empirical data analysis. 
\end{abstract}

\maketitle

\section{Introduction}
Recently, internet trading has become very popular. Obviously, 
the rate (or price) change of the trading behaves according to 
some unknown stochastic processes,  
and numerous studies have 
been conducted to reveal the statistical properties of 
its nonlinear dynamics \cite{Mantegna2000,Bouchaud,Voit}. 
In fact, several authors have analysed 
tick-by-tick data of price changes including 
the currency exchange rate in financial markets 
\cite{Simonsen,Simonsen2,Raberto,Scalas,Kurihara,Sazuka,Sazuka2}. 
Some of these studies are restricted to 
the stochastic variables of price changes 
(returns) and most of them are specified by 
terms such as the {\it fat} or {\it heavy tails} of 
distributions \cite{Mantegna2000}. However, fluctuation in time intervals, 
namely, the duration in the point process \cite{Cox1,Cox2} 
might also contain important market information, 
and it is worthwhile to investigate these properties. 

Such fluctuations in the time intervals between events 
are not unique to price changes 
in financial markets but are also very common in the real world. 
In fact, it is wellknown that the spike train 
of a single neuron in the human brain 
is regarded as a time series, wherein the difference 
between two successive spikes is not constant but fluctuates. 
The stochastic process specified by the so-called 
inter-spike intervals (ISI) 
is one such example \cite{Tuckwell,Gerstner}. 
The average ISI is of the order of a few milli-second 
and the distribution of the intervals is 
well-described by a {\it Gamma distribution} \cite{Gerstner}. 

On the other hand, in financial markets, 
for instance, the time interval between two consecutive 
transactions of Bund futures (Bund is the German word for bond)
and BTP futures
(BTPs are middle- and long-term Italian Government bonds 
with fixed interest rates)
traded at London International 
Financial Futures and Options Exchange (LIFFE) 
is $\sim 10$ seconds 
and is well-fitted by the {\it Mittag-Leffler 
function} \cite{Mainardi,Raberto,Scalas}. 
The Mittag-Leffler 
function behaves as a stretched exponential 
distribution for 
short time-interval regimes, 
whereas for the long time-interval regimes, 
the function has a power-law tail. 
Thus, the behaviour of the distribution 
described by the Mittag-Leffler function 
changes from the stretched 
exponential to the power-law at some critical point 
\cite{Gorenflo}. 
However, it is nontrivial to 
determine whether the Mittag-Leffler function supports 
any other kind of market data, e.g.
the market data filtered by a rate window. 

Similar to the stochastic processes of price change 
in financial markets, 
the US dollar/Japanese yen (USD/JPY)  
exchange rate of 
Sony Bank \cite{Sony}, 
which is an internet-based bank, 
reproduces its rate by using a 
{\it rate window} with a width of $0.1$ yen for 
its individual customers in Japan. 
That is, if the USD/JPY market rate changes by more than $0.1$ yen, 
the USD/JPY Sony Bank rate is updated to the market rate. 
In this sense, it is possible for us to say that the procedure 
of determination of the Sony Bank USD/JPY exchange rate 
is essentially a first-passage process 
\cite{Redner,Kappen,Risken,Gardiner,Schoutens}.

In this paper, we analyse the average time interval that 
a customer must wait until the next price (rate) change 
after they log in to their computer systems.
Empirical data analysis has shown 
that the average time interval between rate changes is one 
of the most important statistics for understanding market behaviour. 
However, as internet trading becomes popular, 
customers would be more interested in the average waiting time 
--- defined as the average time that customers have to wait 
between any instant and the next price change  --- 
when they want to observe the rate, e.g. 
when they log in to their computer systems 
rather than the average time interval between rate changes. 
To evaluate the average waiting time, we use the 
so-called renewal-reward theorem which is wellknown in 
the field of queueing theory \cite{Tijms,Oishi}. 
In addition, we provide a simple formula 
to evaluate the expected reward for customers. 
In particular, we investigate these important quantities for 
the Sony Bank USD/JPY exchange rate by analysing a 
simple probabilistic model and computer simulations 
that are 
based on the {\it ARCH (autoregressive 
conditional heteroscedasticity)} \cite{Engle} and 
{\it GARCH (generalised ARCH)} \cite{Mantegna2000,Ballerslev,Franke} 
stochastic 
models with the assistance of 
empirical data analysis of 
the Sony Bank rate \cite{Sazuka,Sazuka2}.

This paper is organised as follows. 
In the next section, we explain the 
method being used by 
the Sony Bank and introduce several studies concerning 
empirical data about the rate. In Sec. III, we introduce 
a general formula to evaluate average waiting time 
using the renewal-reward theorem and calculate it 
with regard to Sony Bank customers. 
Recently, one of the authors \cite{Sazuka2} 
provided evidence implying that 
the first-passage time (FPT) distribution of the Sony Bank rate 
obeys the {\it Weibull distribution} \cite{Everitt}. 
This conjecture is regarded as a counter part 
of studies that suggest that the FPT should follow 
an exponential distribution (see example in 
\cite{Chan}). In the same section, 
we evaluate the average waiting time 
while assuming that FPT obeys an exponential 
distribution. Next, 
we compare it with the result for the Weibull distribution and 
perform empirical data analyses \cite{Sazuka,Sazuka2}. 
We find that the assumption of 
exponential distributions on the first-passage process 
should be rejected and that a Weibull distribution seems 
more suitable for 
explaining the first-passage processes of the Sony Bank rate. 
Thus, we can predict how long 
customers wait and how many returns 
they obtain until the next rate change 
after they log in to their computer systems. 
In Sec. IV, 
to investigate the effects of the rate window of 
the Sony Bank, we introduce 
the ARCH and GARCH models to reproduce the raw data 
before filtering it 
through the rate window. 
In Sec. V, we evaluate the reward 
that customers can expect to obtain after they 
log in to their computer systems. 
The last section provides a summary and discussion.

\section{Data: The Sony Bank USD/JPY exchange rate}

\begin{figure}[ht]
\begin{center}
\includegraphics[width=12cm]{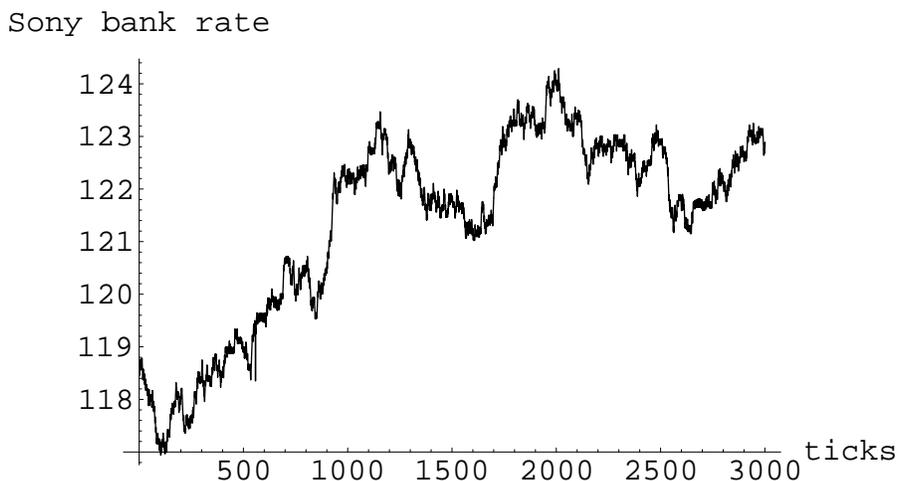}
\end{center}
\caption{\footnotesize 
The Sony Bank USD/JPY exchange rate. 
The units of the horizontal and vertical axes are ticks and 
yen/dollar, respectively.
}
\label{fig:Sazuka3000}
\end{figure}

\begin{figure}[ht]
\begin{center}
\includegraphics[width=10cm]{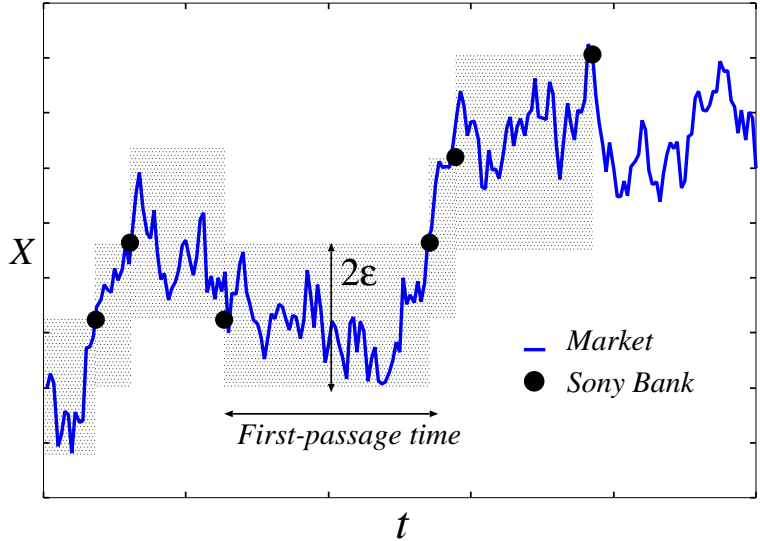}
\end{center}
\caption{\footnotesize An illustration of 
generating the filtered rate 
(black circle) by the 
rate window with 
a constant width $2\epsilon$ in time 
(shaded area) from the 
market rate (solid line). 
The units of the horizontal and the vertical axes are 
ticks and yen, respectively.}
\label{fig:fg_window}
\end{figure}

The Sony Bank rate \cite{Sony} is 
the foreign exchange rate that the 
Sony Bank offers with reference to the market rate and 
not their customers' orders. 
In FIG. \ref{fig:Sazuka3000}, 
we show a typical update of 
the Sony Bank rate. 
If the USD/JPY market rate 
changes by greater than or equal to $0.1$ yen, 
the Sony Bank USD/JPY 
rate is updated to the market rate. 
In FIG. \ref{fig:fg_window}, 
we show the method of generating the Sony Bank rate 
from the market rate. 
In this sense, the Sony Bank rate 
can be regarded as a kind of first-passage processes. 
In TABLE \ref{tab:table1}, 
we show data concerning 
the Sony Bank USD/JPY 
rate vs. tick-by-tick 
data for the USD/JPY rate from Bloomberg L.P. 
\begin{table}
\caption{\label{tab:table1}
The Sony Bank USD/JPY rate vs. tick-by-tick 
data for the USD/JPY rate \cite{Sazuka}. 
 }
\begin{ruledtabular}
\begin{tabular}{lcc}
\mbox{} & Sony Bank rate & tick-by-tick data \\
\hline
Number of data per day & $\sim 70$  & $\sim 10,000$ \\
The smallest price change  & $0.1$ yen  & $0.01$ yen \\
Average interval between data & $\sim 20$ minutes & $\sim 7$ seconds \\
\end{tabular}
\end{ruledtabular}
\end{table}
From these tables and figures, 
an important question might arise. Namely; how long should 
Sony Bank customers wait
between observing the price 
and the next price change? 
This type of 
question never arises 
in the case of ISI or Bund futures, because 
the average time intervals are too short to 
evaluate such measures. 
We would like to stress that in this paper, 
we do not discuss the 
market data underlying the Sony Bank rate; 
however, as we will see in the following 
sections, the main result obtained in 
this study, i.e. 
how long a trader must wait for the next price 
change of the Sony Bank rate 
from the time she or he logs on to the internet, 
is not affected by this lack of information. 
\section{Renewal-reward theorem and average waiting time}
From TABLE \ref{tab:table1}, 
we find that 
the number of 
data per day is remarkably 
reduced from 
$\sim 10,000$ to 
$\sim 70$ because of an effect of the rate window 
of $0.1$-yen width. 
As a result, the average interval of 
exchange rate updates is extended to $\sim 20$ min. 
This quantity is one of the 
most important measures for the market, 
however, 
customers 
might seek information about 
the {\it average waiting time}, 
which 
is defined 
as the average 
time interval 
that customers 
have to wait 
until the next 
change of the Sony Bank USD/JPY rate 
after they 
log in to their computer systems. 
To evaluate 
the average waiting time, 
we use the {\it renewal-reward theorem}, which 
is wellknown in the field 
of queueing theory \cite{Tijms,Oishi}.
We briefly explain the theorem below. 

Let us define 
$N(\tau)$ as the number of 
rate changes within 
the interval $(0,\tau]$ and 
suppose that a customer 
logs in to her/his computer system 
at time $\tau$ 
($\tau_{N(\tau)} \leq 
\tau < \tau_{N(\tau)+1}$). 
Then, 
we defined 
the following quantity: 
\begin{eqnarray}
W(\tau) & = & 
\tau_{N(\tau)+1}-\tau, 
\end{eqnarray}
which denotes 
the waiting time 
for the customer until 
the next update of the 
Sony Bank USD/JPY rates 
when she or he logs in to a computer system 
at time $\tau$. 
Then, 
the renewal-reward theorem \cite{Tijms,Oishi} 
implies that 
the average 
waiting time $w$ 
can be written in terms of 
the first two moments, 
$E(t)$ and $E(t^{2})$, of the first-passage time 
distribution $P(t)$ as 
\begin{eqnarray}
w & = & 
\lim_{\tau \to \infty}
\frac{1}{\tau}
\int_{0}^{\tau}
W(s)ds = 
\frac{E(t^{2})}
{2E(t)}, 
\label{eq:RR}
\end{eqnarray}
where 
$E(\cdots)$ denotes 
$\int_{0}^{\infty}dt (\cdots) P(t)$. 
Therefore, if 
we obtain the explicit form of 
the FPT distribution $P(t)$, 
we can evaluate 
the average waiting time $w$ 
by using this theorem (\ref{eq:RR}). 

The proof of the theorem 
is quite simple. 
Let us suppose that the points at which 
the Sony Bank USD/JPY rate  
changes are given by 
the time sequence 
$\tau_{1},\tau_{2},\cdots, 
\tau_{N(\tau)}$. 
For these data 
points, 
the first-passage time series is given by 
definition as 
$t_{0}=\tau_{1}, t_{1}=\tau_{2}-\tau_{1}, 
t_{2}=\tau_{3}-\tau_{2}, 
\cdots, 
t_{N(\tau)-1}=
\tau_{N(\tau)}-\tau_{N(\tau)-1}$. 
Then, 
we observe that 
the time 
integral 
$\int_{0}^{\tau}
W(s)ds$ appearing 
in equation (\ref{eq:RR}) 
is identical to 
$\sum_{i=0}^{N(\tau)}(t_{i})^{2}/2$, 
where 
$(t_{i})^{2}/2$ 
corresponds to 
the area of a 
triangle with sides 
$t_{i},t_{i}$ and 
$\sqrt{2}\,t_{i}$. 
As a result, we obtain 
\begin{eqnarray}
w & = & 
\lim_{\tau \to \infty}
\frac{1}{\tau}
\int_{0}^{\tau}
W(s)ds  \,\simeq \,
\frac{N(\tau)}{\tau}
\cdot 
\frac{1}{N(\tau)}
\sum_{i=1}^{N(\tau)}
\frac{(t_{i})^{2}}{2} = 
\frac{E(t^{2})}{2E(t)},
\end{eqnarray}
wherein 
we used 
the fact that 
the expectation of 
the waiting time $t_{i}$ 
is given by 
\begin{eqnarray}
E(t) & = & 
\frac{1}{N(\tau)}
\sum_{i=0}^{N(\tau)}
t_{i}=
\frac{1}{N(\tau)}
\{
\tau_{1}+
\tau_{2}-\tau_{1}+
\tau_{3}-\tau_{2}+
\cdots + 
\tau_{N(\tau)}
-\tau_{N(\tau)-1}
\}=
\frac{\tau_{N(\tau)}}
{N(\tau)}.
\end{eqnarray}
Therefore,  
\begin{eqnarray}
\lim_{\tau \to \infty}
\frac{N(\tau)}{\tau} & = & 
\frac{1}{E(t)}.
\end{eqnarray}
Thus, 
equation (\ref{eq:RR}) holds true. 

Therefore, 
if we obtain the explicit form of the 
first-passage time distribution $P(t)$, the average 
waiting time $w$ can be evaluated by theorem 
(\ref{eq:RR}). 
However, for estimating 
the distribution 
$P(t)$, we can only perform empirical data analysis of 
the Sony Bank rate. 
Apparently, the first-passage process of the Sony Bank rate 
depends on the stochastic processes 
of raw market data underlying the data that is available on the 
website for customers 
after being filtered by a rate window of $0.1$-yen width. 
Unfortunately, 
this raw data is not available to us 
and we cannot use 
any information regarding the high-frequency raw data.

Recently, an empirical data analysis 
by one of the authors of this paper \cite{Sazuka2} 
revealed that the 
Weibull probability distribution function, which is defined by 
\begin{eqnarray}
P_{W}(t) & = & 
m\,
\frac{t^{m-1}}{a}\,
{\exp}
\left(
-\frac{t^{m}}{a}
\right), 
\label{eq:Weibull}
\end{eqnarray}
is a good approximation of 
the Sony Bank USD/JPY rate in the non-asymptotic regime. 
We should keep in mind 
that in the asymptotic regime, 
the FPT distribution 
obeys the power-law --- $P(t) \,\sim \,
t^{-\alpha}\,\,
(\alpha \sim 3.67)$ --- and 
a crossover takes place at the 
intermediate 
time scale $t_{\rm c} \,\sim\, 18,000$ [s]. 
However, 
the percentage of empirical data 
for the power-law regime is 
only $0.3 \%$. 
Similar behaviour was observed 
for the time-interval distribution 
between two consecutive 
transactions of Bund futures 
traded at LIFFE, and the time-interval 
distribution is well-described by  
the Mittag-Leffler type 
function \cite{Mainardi}, 
namely, 
the distribution changes from the stretched exponential to 
the power-law for long-time regimes. 
Therefore, 
we have chosen a Weibull distribution 
in both asymptotic and non-asymptotic 
regimes to evaluate the average waiting time. 
The justification for this choice 
will be discussed later 
by comparing 
the predicted value with 
the result from empirical data analysis. 

This Weibull distribution (\ref{eq:Weibull}) 
is reduced to an exponential distribution 
for 
$m=1$ and a Rayleigh distribution for 
$m=2$. 
It is easy for us to 
evaluate the first two moments of 
the distribution $P_{W}(t)$. 
We obtained them explicitly as 
$E(t)=a^{1/m}\Gamma(1+(1/m))$ and 
$E(t^{2})=a^{2/m}
\Gamma(1+(2/m))$. 
Thus, the average waiting time $w$
for the Weibull distribution is 
explicitly given by  
\begin{eqnarray}
w & = & 
a^{1/m}\,
\frac{
\Gamma
\left(
\frac{2}{m}
\right)
}
{
\Gamma
\left(
\frac{1}{m}
\right)
}, 
\label{eq:w_Weibull}
\end{eqnarray}
where $\Gamma (z)$ refers to the gamma function, and 
$\Gamma(z+1)=z\Gamma(z)$. 
\begin{figure}[ht]
\begin{center}
\rotatebox{-90}{\includegraphics[width=8cm]{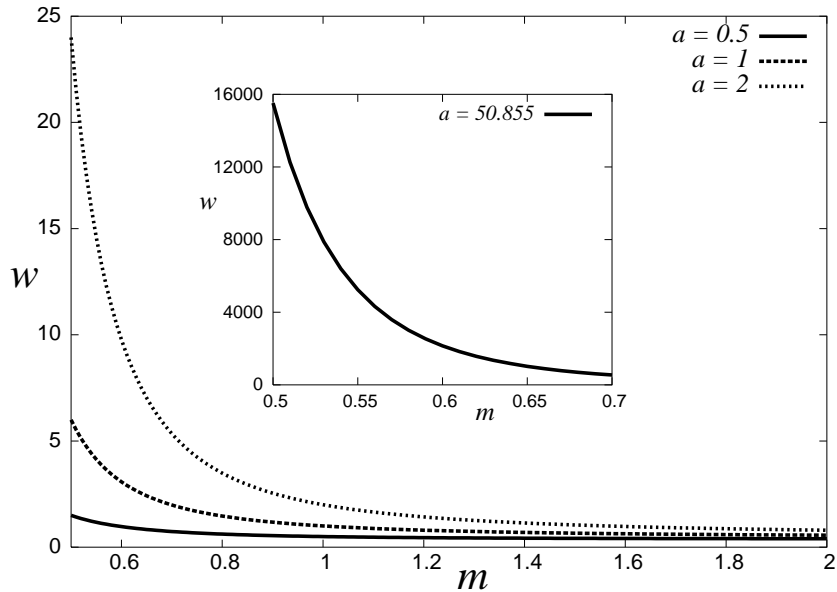}}
\end{center}
\caption{\footnotesize 
Average 
waiting time $w$ as a function of 
$m$ for the Weibull distribution. 
We set $a=0.5,1$ and $2$. 
The unit of the vertical axis is second. 
In the inset, 
we plot for 
$a=50.855$, which was 
evaluated from empirical 
data of the Sony Bank 
USD/JPY rate \cite{Sazuka2}. 
From the inset, 
for the parameter $m = 0.59$,  
the average waiting time $w$ 
is expected to be $w \simeq 2534.146
 \,\,\mbox{[s]} =42.236$ [min]. 
This time interval 
does not differ much from 
the result ($w \sim 49$ [min]) 
obtained from our empirical data analysis. 
}
\label{fig:fg2}
\end{figure}
In FIG. \ref{fig:fg2} we plot the 
average waiting time $w$ (\ref{eq:w_Weibull}) 
as a function 
of $m$ for 
several values of $a$, and we find that 
$w$ for Sony Bank customers is 
convenient and reasonable. 
This is 
because 
the empirical data analysis 
reported the value of the parameter as 
$m =0.59$, and 
the average waiting time $w$ for 
this case is evaluated as $w \simeq 2534.146 \,\,\mbox{[s]} 
= 42.236$ [min] from the inset of this figure. 
From reference \cite{Sazuka2}, 
information about the scaling parameter 
$a$ was not available; however, 
we can easily obtain the value as discussed next. 

As mentioned above, 
several empirical data analyses \cite{Sazuka2} 
revealed that parameter $m \simeq 0.59$ 
for the Sony Bank rate. 
In fact, it is possible for us 
to estimate 
$a$ by using 
the fact that 
the average interval $\langle t \rangle$ 
of the rate change is 
$\sim 20$ [min] (TABLE \ref{tab:table1}). 
Next, 
we obtain 
the following 
simple relation: 
\begin{eqnarray}
E(t) & = & 
\langle t \rangle.
\end{eqnarray}
That is 
$(a^{1/m}/{m})\Gamma(1/m)=\langle t \rangle$. 
Therefore, 
\begin{eqnarray}
a & = & 
\left\{
\frac{m \langle t \rangle}
{\Gamma 
\left(
\frac{1}{m}
\right)
}
\right\}^{m}.
\end{eqnarray}
Substituting 
$m \simeq 0.59$ and 
$\langle 
t \rangle \sim 
20 \times 60 = 1200$ [s], 
we obtain the parameter 
$a$ for the Sony Bank rate 
as $a \simeq 50.855$.

It should be noted that  
the average waiting time 
is also evaluated by a simple sampling 
from the empirical data of the Sony Bank rates. 
The first two moments 
of the first-passage time distribution 
$P_{W}(t)$ --- 
$\langle t \rangle \equiv 
\lim_{T \to \infty}
(1/T) \sum_{k=0}^{T} \Delta t_{k}, 
\langle t^{2} \rangle \equiv 
\lim_{T \to \infty}
(1/T) \sum_{k=0}^{T}
(\Delta t_{k})^{2}$ --- are easily calculated 
from the sampling, and then, 
the average waiting time 
is given by 
$w_{\rm sampling} = 
\langle t^{2} \rangle /
2\langle t \rangle$. 
We find that $w_{\rm sampling} \simeq 49$ [min] 
which 
is not much different from that obtained by 
evaluation ($w=42.236$ [min]) 
by means of the renewal-reward theorem (\ref{eq:RR}). 
Thus, the renewal-reward theorem introduced here 
determines the average waiting time of the 
financial system with adequate accuracy, and our assumption 
of the Weibull distribution 
for the first-passage time of the Sony Bank rates 
seems to be reasonable. 
We would particularly like to 
stress that 
the $0.3\%$ 
data from the asymptotic 
regime does not contribute much to 
the average waiting time. 

A detailed account of this fine agreement 
between $w$ and $w_{\rm sampling}$, 
more precisely, 
the factors to which we can 
attribute the small difference between 
$w$ and $w_{\rm sampling}$, will 
be reported in our forthcoming paper \cite{ISE}. 

We would now like to discuss 
the average waiting time 
for the case in which 
the first-passage process 
can be regarded as a Poisson process. 
In the Poisson process, 
the number of 
events occurring in the interval 
$(0,\tau]$ 
is given by $P_{k}(\tau)=(\lambda \tau)^{k}\, 
{\rm e}^{-\lambda \tau}
/k!$\,\,\,$(k=0,1,2,\cdots)$. 
For this Poisson 
process, 
the time 
interval $t$ between two arbitrary successive 
events obeys the following 
exponential 
distribution: 
\begin{eqnarray}
P(t) & = & 
\lambda \,
{\rm e}^{-\lambda t}.
\end{eqnarray}
Then, we obtain 
the first two moments of the 
above distribution 
$P(t)$ as 
$E(t)=1/\lambda, 
E(t^{2})=2/\lambda^{2}$. 
These two moments 
give the 
average waiting time 
for the Poisson 
process as  
$w=E(t^{2})/2E(t)=1/\lambda = E(t)$. 
Therefore, 
for the Poisson 
process, 
the average 
waiting time $w$ 
is identical to 
the average rate interval 
$E(t)$. 
This result might be naturally 
accepted because the login-time 
dependence on the average waiting time 
is averaged out due to 
the fact that for a Poisson process, 
each event takes place independently. 
On the other hand, 
for a non-Poisson process, the login time 
is essential to determine the average waiting time. 
Thus, the average waiting time 
becomes $E(t^{2})/2E(t)$ 
instead of $E(t)$. 

We should stress that for a given stochastic process 
whose FPT is a constant $T$, 
the average waiting time 
$w$ is easily evaluated by 
$w=(1/T)\int_{0}^{T}s ds=T/2$. 
However, 
the result for the above 
Poisson 
process implies that 
for a Poisson process with a mean $E(t)$ of the FPT, 
the average waiting time 
is not $E(t)/2$ but $E(t)$. 
This fact is referred to as 
the {\it inspection paradox} \cite{Tijms} 
in the field of queueing theory. 
For a Poisson 
process, each event occurs independently, 
but the time interval between 
the events follows an exponential 
distribution. 
Therefore, 
there is some opportunity for the customers to 
log in to their computer systems 
when the exchange rate remains the same  
for a relatively 
long time, 
although the short FPT occurs with higher 
frequency 
than the long FPT. 
This fact makes 
$w$ the first moment $E(t)$ of 
the FPT distribution. 

On the other hand, for the Weibull 
distribution, 
the condition 
under which the average waiting time 
$w$ is the same as 
the average first-passage time 
$E(t)$ is given by $w=E(t)$, i.e. 
\begin{eqnarray}
m \Gamma
\left(
\frac{2}{m}
\right) & = & 
\left\{
\Gamma
\left(
\frac{1}{m}
\right)
\right\}^{2}.
\end{eqnarray}
\begin{figure}[ht]
\begin{center}
\rotatebox{-90}{\includegraphics[width=8cm]{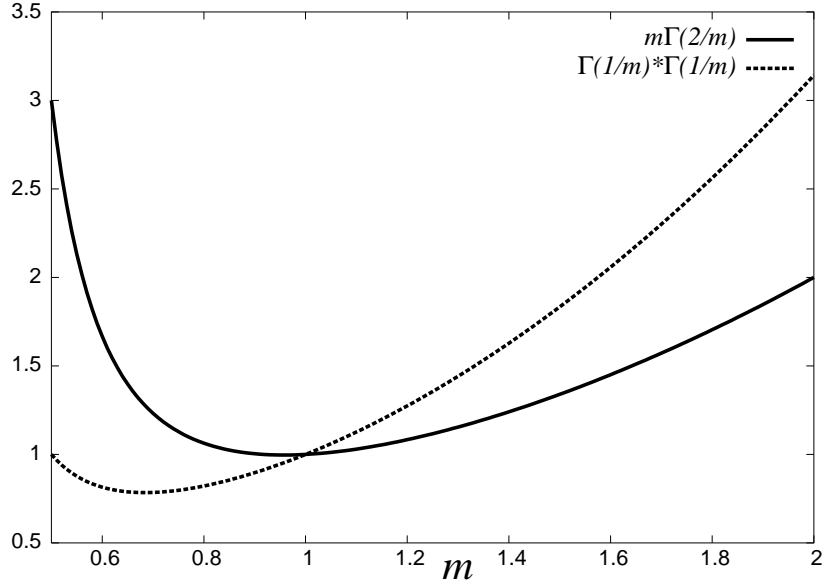}}
\end{center}
\caption{\footnotesize 
$m\Gamma(2/m)$ and 
$\{\Gamma(1/m)\}^{2}$ 
as functions of 
$m$. 
The crossing point $m_{\times}=1$ of the 
two curves 
corresponds to 
the Weibull parameter for 
which 
the average waiting time 
$w$ is 
identical to the average 
first-passage time $E(t)$. 
For $m < m_{\times}$ 
(the case of Sony Bank), 
$w > E(t)$ holds true, 
whereas for 
$m > m_{\times}$, $w < E(t)$ is satisfied.}
\label{fig:fg4}
\end{figure}
In FIG. \ref{fig:fg4}, 
we plot both curves 
$m\Gamma(2/m)$ and 
$\{\Gamma(1/m)\}^{2}$,  
as functions of $m$. 
The crossing point $m_{\times}=1$ 
corresponds to 
the Weibull parameter 
for which 
the average waiting time 
$w$ is 
identical to the average 
first-passage time $E(t)$. 
Obviously, 
for the parameter 
regime $m < m_{\times}$ (the case of Sony Bank), 
the average waiting time evaluated by 
the renewal-reward 
theorem becomes longer than 
the average first-passage time $E(t)$. 
As mentioned before, 
the Weibull distribution with 
$m=1$ follows the exponential 
distribution $P_{W}(t)\,\sim\, 
{\rm e}^{-t}$. 
Therefore, 
as mentioned in the case of 
the Poisson process, 
the login-time 
dependence of 
the customers is 
averaged out in the calculation 
of the average waiting time $w$.  
As the result, 
$w$ becomes 
the first moment of 
the FPT distribution, $E(t)$. 
However, once the parameter of the 
Weibull distribution has deviated from $m=m_{\times}=1$, 
the cumulative distribution is no longer an exponential 
distribution. 
Thus, the stochastic process 
is not described 
by a Poisson 
process and the 
average 
waiting time $w$ is not 
$E(t)$ but 
$w=E(t^{2})/2E(t)$ which is 
derived 
from the renewal-reward 
theorem. 
As mentioned above, 
the empirical data 
analysis 
revealed 
that 
the average waiting time 
of Sony Bank rates 
is $w_{sampling}=\langle t^{2} \rangle/2\langle t \rangle \sim 49$ [min]. 
This value is 
almost twice of 
the average interval 
of the rate change 
data $E(t)=\langle t \rangle \,\sim\, 20$ [min] 
in TABLE \ref{tab:table1}. 
This fact is 
a kind of justification or 
evidence for the conclusion that 
the Sony Bank USD/JPY exchange rate obeys 
non-exponential 
distribution 
from the viewpoint of the renewal-reward theorem. 
\section{Effect of rate window on first-passage processes 
of price change}
In the previous sections, 
we evaluated the average waiting time 
$w$ for foreign exchange rates (price change). 
In particular, we focused on the 
average waiting time of the 
Sony Bank rate whose 
FPT distribution 
might be represented by a Weibull distribution. 
We compared 
the average waiting time 
$w$ obtained by the renewal-reward theorem with 
that obtained by a simple sampling of 
the empirical data of the Sony Bank rate and 
found a good agreement between the two. 
However, as mentioned earlier, 
the empirical data of Sony Bank was obtained 
by filtering the raw market data 
using a rate window of $0.1$-yen width. 
Unfortunately, 
the raw data 
of the real market 
is not available to us 
(Sony Bank does not keep these records) and 
the only information we could obtain is 
the Sony Bank rate itself. 
Although the main result of the evaluation of the average waiting 
time is never affected by this lack of information, 
it might be worthwhile to investigate the role of 
the rate window with the assistance of 
computer simulations. 

For this purpose, 
we introduce the stochastic process 
$X_{k_{0}},
X_{k_{1}},\cdots, 
X_{k_{n}},\cdots$ , 
where the time intervals of two successive time stamps 
$\Delta t_{1}=k_{1}-k_{0}, 
\Delta t_{2}=k_{2}-k_{1}, 
\cdots, \Delta t_{n}=k_{n}-k_{n-1}, \cdots$ 
obey a Weibull distribution 
specified by the two parameters $m_{0}$ and 
$a_{0}$. 
For example, 
for the ordinary Wiener process, 
the above 
stochastic process 
is described by 
\begin{eqnarray}
X_{t+\Delta t} & = & X_{t} +  
Y_{t},\,\,\,\,\, Y_{t}={\cal N}(0,\sigma^{2}), 
\label{eq:Wiener2} \\
P_{W}^{0}(\Delta t) & = & 
m_{0}\,
\frac{(\Delta t)^{m_{0}-1}}{a_{0}}
{\exp}
\left(
-\frac{(\Delta t)^{m_{0}}}{a_{0}}
\right),
\label{eq:Weibull_m0}
\end{eqnarray}
where $X_{t}$ denotes 
the value of price at time $t$. 
Obviously, 
the first-passage time of the above stochastic 
process has a distribution $P(t)$, 
which is generally different from $P_{W}^{0} (\Delta t)$. 
Nevertheless, here we assume that 
$P(t)$ also obeys a Weibull distribution 
with parameters $m$ and $a$. 
Next, 
we investigate 
the effect of a rate 
window with a 
width of $\epsilon$ through 
the difference of 
the parameters, 
namely, 
the $m$-$m_{0}$ and 
$a$-$a_{0}$ plots. 
For simplicity, 
we set $a=a_{0}=1$ in 
our computer simulations. 
In other words, 
using computer simulations, 
we clarify 
the relationship 
between the {\it input} 
$m_{0}$ and 
the corresponding 
{\it output} $m$ 
of the filter with a rate 
window of width $\epsilon$. 
\subsection{The Weibull paper analysis of the FPT distribution}
To determine the output $m$ from 
the histogram of first-passage time, 
we use 
the {\it Weibull 
paper 
analysis} \cite{Sazuka2, Everitt} 
for cumulative 
distribution. 
Now, we briefly 
explain the details of the method. 

We first consider the 
cumulative distribution of 
the Weibull distributions with $a=1$ as  
\begin{eqnarray}
R (t) & = & 
\int_{t}^{\infty} 
P_{W}(s)ds = \exp (-t^{m}).
\end{eqnarray}
Then, for the 
histogram of the cumulative distribution 
obtained by sampling using 
computer simulations, 
we fit the histogram to the 
Weibull distribution 
by using 
the following 
{\it Weibull paper}: 
\begin{eqnarray}
\log \{\log 
(R(t)^{-1})\} & = & 
m\log t.
\label{eq:WeibullP}
\end{eqnarray}
Thus, 
the parameter $m$ of the 
first-passage time 
distribution $P(t)$ 
is obtained as a slope of 
the $\log t$-$\log \log 
(R(t)^{-1})$ plot. 
In the following subsection, 
we evaluate 
the parameter $m$ of 
the FPT distribution for 
the stochastic processes 
with time intervals of 
two successive time stamps $\Delta t$ 
obeying the Weibull distribution $P_{W}^{0} (\Delta t)$. 
Then, we compare $m_{0}$ with $m$ using the above Weibull 
paper analysis (\ref{eq:WeibullP}). 
\subsection{ARCH and GARCH processes as 
stochastic models for raw market data}
As stochastic model 
of the raw data of the real market, the 
Wiener process defined by equations 
(\ref{eq:Wiener2}) and (\ref{eq:Weibull_m0}) is 
one of the simplest candidates. 
However, numerous studies 
from both 
empirical and theoretical viewpoints 
have revealed that 
the volatility  
of financial data such as the USD/JPY exchange rate 
is a time-dependent stochastic variable. 
With this fact in mind, 
we introduce 
two types of stochastic 
processes that are 
characterised by time-dependent volatility --- 
{\it ARCH} \cite{Engle} and 
{\it GARCH} models \cite{Mantegna2000,Engle,Franke}.

The ARCH(1) model used in this section 
is described as follows: 
\begin{eqnarray}
\sigma_{t+\Delta t}^{2} & = & 
\alpha_{0} + 
\alpha_{1} X_{t}^{2}, 
\label{eq:ARCH1} \\
X_{t +\Delta t} & = & X_{t} + 
{\cal N}(0,\sigma_{t}^{2}),
\label{eq:ARCH2}
\end{eqnarray}
where we assume that the time interval $\Delta t$ 
obeys the Weibull distribution 
(\ref{eq:Weibull_m0}). 
From the definition in 
(\ref{eq:ARCH1}) and (\ref{eq:ARCH2}), 
we find that the volatility $\sigma_{t}^{2}$ 
controlling 
the conditional 
probability density $P(X_{t})$ 
at time $t$ fluctuates. 
However,
for the limit $t \to \infty$, 
such a {\it local time dependence} does not 
prevent the stochastic 
process 
from having a well-defined 
asymptotic 
distribution 
$P(X)$. 
In fact, the above ARCH(1) model 
is characterised 
by the variance $\sigma^{2}$ 
observed over a long time interval 
$t \to \infty$. 
It is easily shown that 
$\sigma^{2}$ can be written in terms of 
$\alpha_{0}$ and $\alpha_{1}$ as 
\begin{eqnarray}
\sigma^{2} & = & 
\frac{\alpha_{0}}
{1-\alpha_{1}}.
\end{eqnarray}
We choose the parameters $\alpha_{0}$ and 
$\alpha_{1}$ so as to satisfy 
$\sigma=\epsilon=1$. 
As a possible choice for 
this requirement, 
we select $(\alpha_{0},\alpha_{1})=(0.45,0.55)$. 
For this parameter choice, 
the {\it kurtosis} (${\rm Kurt}$), 
which 
is defined by the second and fourth 
moments of the probability 
distribution function 
of the stochastic variable 
$X$ as 
$\langle X^{4} \rangle/
\langle X^{2} \rangle^{2}$, leads to 
\begin{eqnarray}
{\rm Kurt} & = & 
3 + 
\frac{6\alpha_{1}^{2}}
{1-3\alpha_{1}^{2}}.
\end{eqnarray}
As is well-known, 
the higher kurtosis values of 
financial 
data indicate that 
the values close to 
the mean and extreme positive and 
negative outliers appear 
more frequently than for 
normally distributed variables. 
In other words, 
the kurtosis is a measure of 
the fatness of the tails 
of the distribution. 
For instance, 
a normal Gaussian distribution has 
${\rm Kurt}=0$, 
whereas distributions with 
${\rm Kurt} >0$ 
are referred to as 
{\it leptokurtic} and 
have tails fatter than 
a Gaussian distribution. 
The kurtosis 
of the ARCH(1) model with 
$(\alpha_{0},\alpha_{1})=(0.45,0.55)$ 
is ${\rm Kurt}=23$. 

Here, we also introduce 
the GARCH(1,1) model defined as 
\begin{eqnarray}
\sigma_{t+\Delta t}^{2} & = & 
\alpha_{0} + 
\alpha_{1}X_{t}^{2} + 
\beta_{1} \sigma_{t}^{2}, \\
X_{t+\Delta t} & = & 
X_{t} + 
{\cal N}(0,\sigma_{t}^{2}),
\end{eqnarray}
where the time 
interval $\Delta t$ is 
assumed to obey 
the Weibull distribution 
(\ref{eq:Weibull_m0}). 
The variance of the above 
GARCH(1,1) 
model observed on a long time 
interval, $t \to \infty$, 
is given by 
\begin{eqnarray}
\sigma^{2} & = & 
\frac{\alpha_{0}}
{1-\alpha_{1}-\beta_{1}}.
\end{eqnarray}
To compare the effect of 
the rate window for 
the ARCH(1) model with that of 
the GARCH(1,1) model, 
we choose 
the parameters 
$\alpha_{0},\alpha_{1}$ and 
$\beta_{1}$ so as to 
satisfy 
$\sigma=\epsilon=1$. 
Among many candidates, 
we select 
$(\alpha_{0},\alpha_{1},\beta_{1})=
(0.4,0.3,0.3)$, 
which 
gives the kurtosis 
\begin{eqnarray}
{\rm Kurt} & = & 
3 + 
\frac{6\alpha_{1}^{2}}
{1-3\alpha_{1}^{2} -
2\alpha_{1}\beta_{1} -
\beta_{1}^{2}} = 
4.17.
\end{eqnarray}
\begin{figure}[ht]
\begin{center}
\rotatebox{-90}{\includegraphics[width=5.6cm]{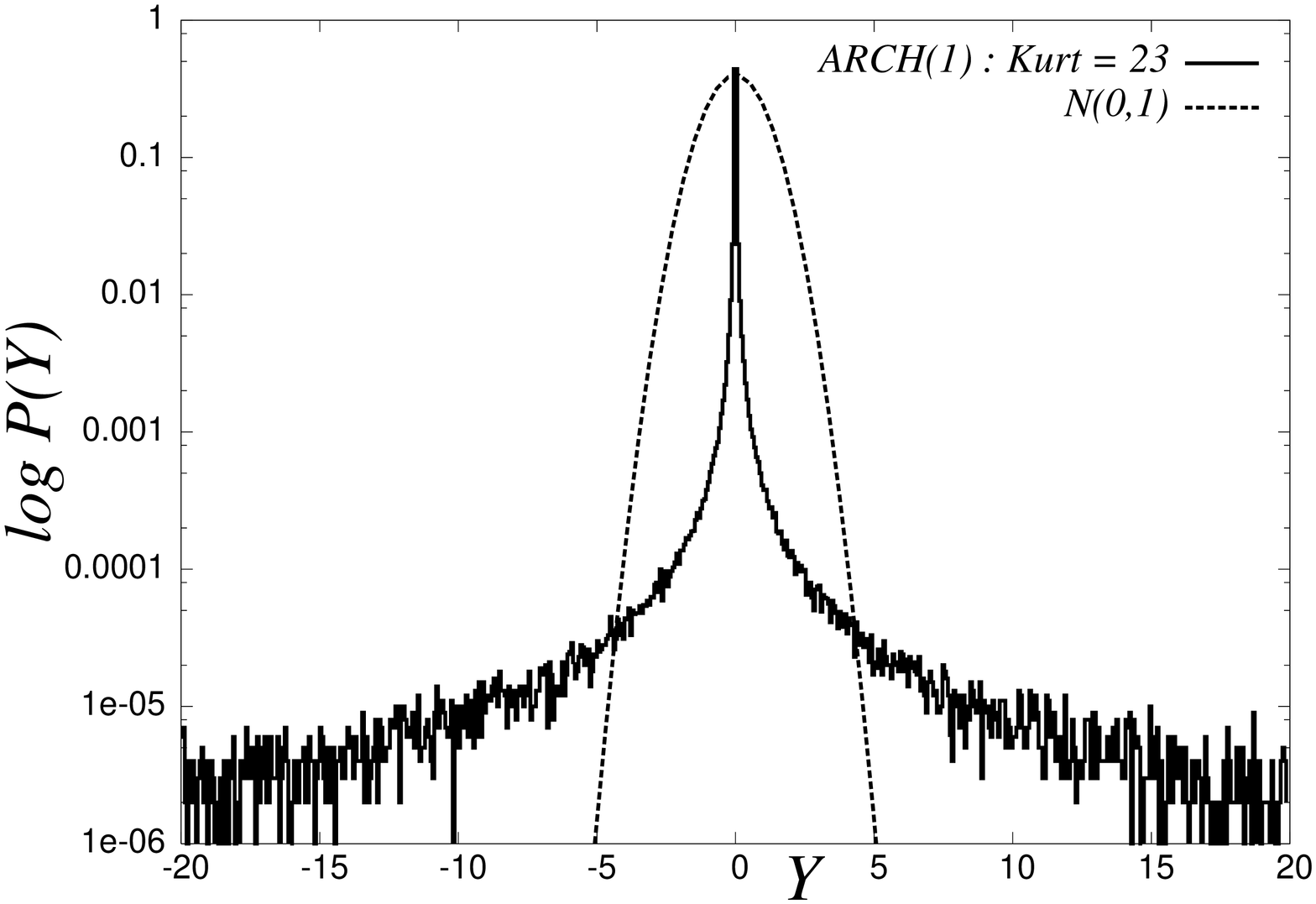}}
\rotatebox{-90}{\includegraphics[width=5.6cm]{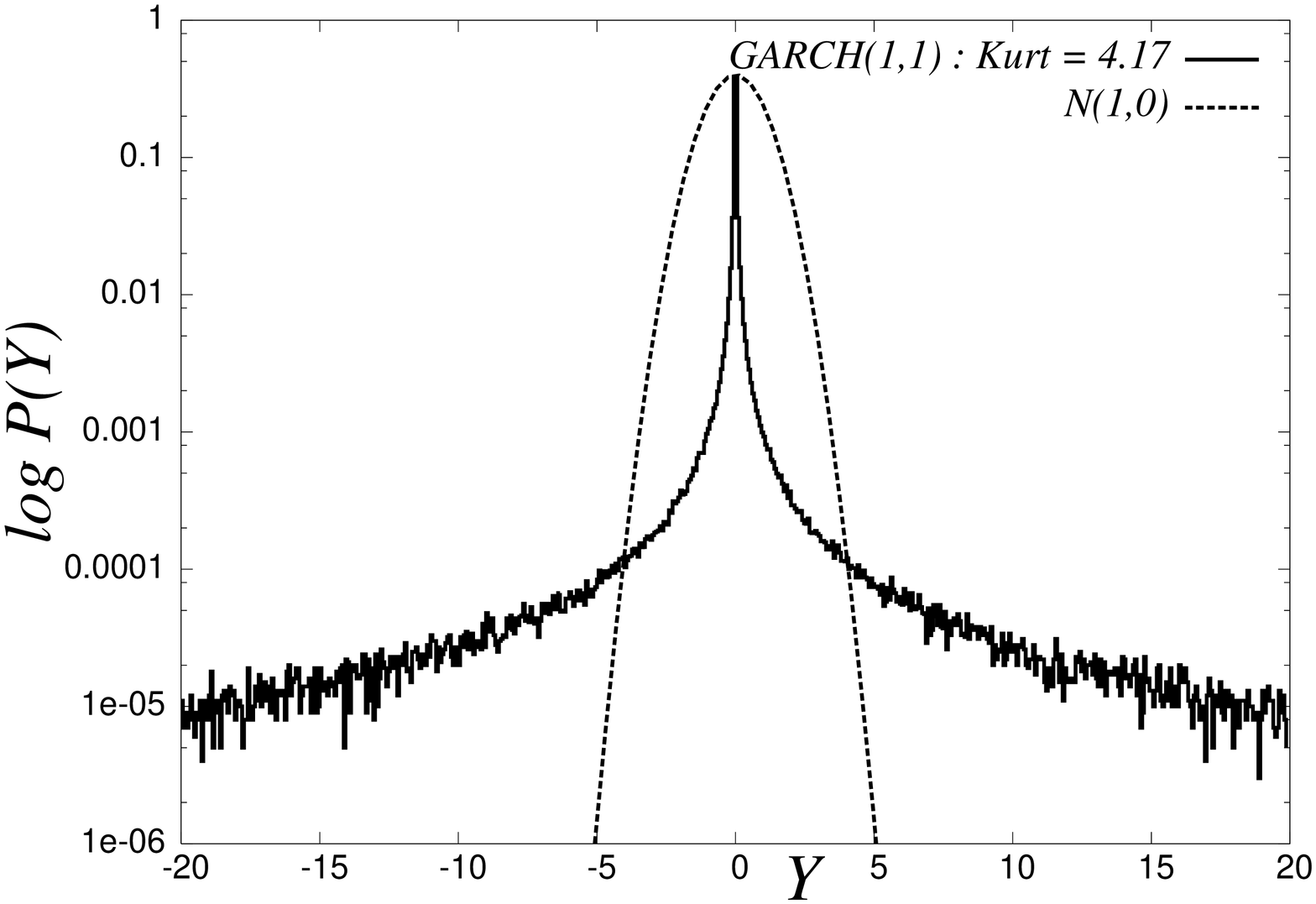}}
\end{center}
\caption{\footnotesize 
Probability density function 
of successive increments (return) 
$Y=X_{t+\Delta t}-X_{t}$. 
The unit of the horizontal axis is yen.
The left panel is the PDF 
for the ARCH(1) model 
with $(\alpha_{0},\alpha_{1})=(0.45,0.55)$ 
and ${\rm Kurt}=23$. 
The right panel is 
the PDF for the GARCH(1,1) model with 
$(\alpha_{0},\alpha_{1},\beta_{1})=
(0.4,0.3,0.3)$ and ${\rm Kurt}=4.17$. 
In both panels, 
the broken line 
corresponds to 
the PDF for a normal Gaussian 
${\cal N}(0,1)$. 
}
\label{fig:fg5}
\end{figure}
In FIG. \ref{fig:fg5}, 
We plot the probability density function (PDF) 
of the successive increments (returns) 
$Y=X_{t+\Delta t}-X_{t}$ for the ARCH(1) and the GARCH(1,1) models. 
The left panel is the PDF 
for the ARCH(1) with $(\alpha_{0},\alpha_{1})=(0.45,0.55)$ 
and ${\rm Kurt}=23$. 
The right panel is 
the PDF for the GARCH(1,1) model 
with 
$(\alpha_{0},\alpha_{1},\beta_{1})=
(0.4,0.3,0.3)$ and ${\rm Kurt}=4.17$.

For the stochastic processes 
for modelling the raw real market data, 
namely, 
the Winer process and 
the ARCH(1) and GARCH(1,1) models, 
we determine 
the parameter $m$ for each 
FPT distribution by means of 
the Weibull paper analysis 
based on (\ref{eq:WeibullP}), 
and then plot the $m_{0}$-$m$ relation 
for each 
stochastic model. 
\begin{figure}[ht]
\begin{center}
\rotatebox{-90}{\includegraphics[width=5.6cm]{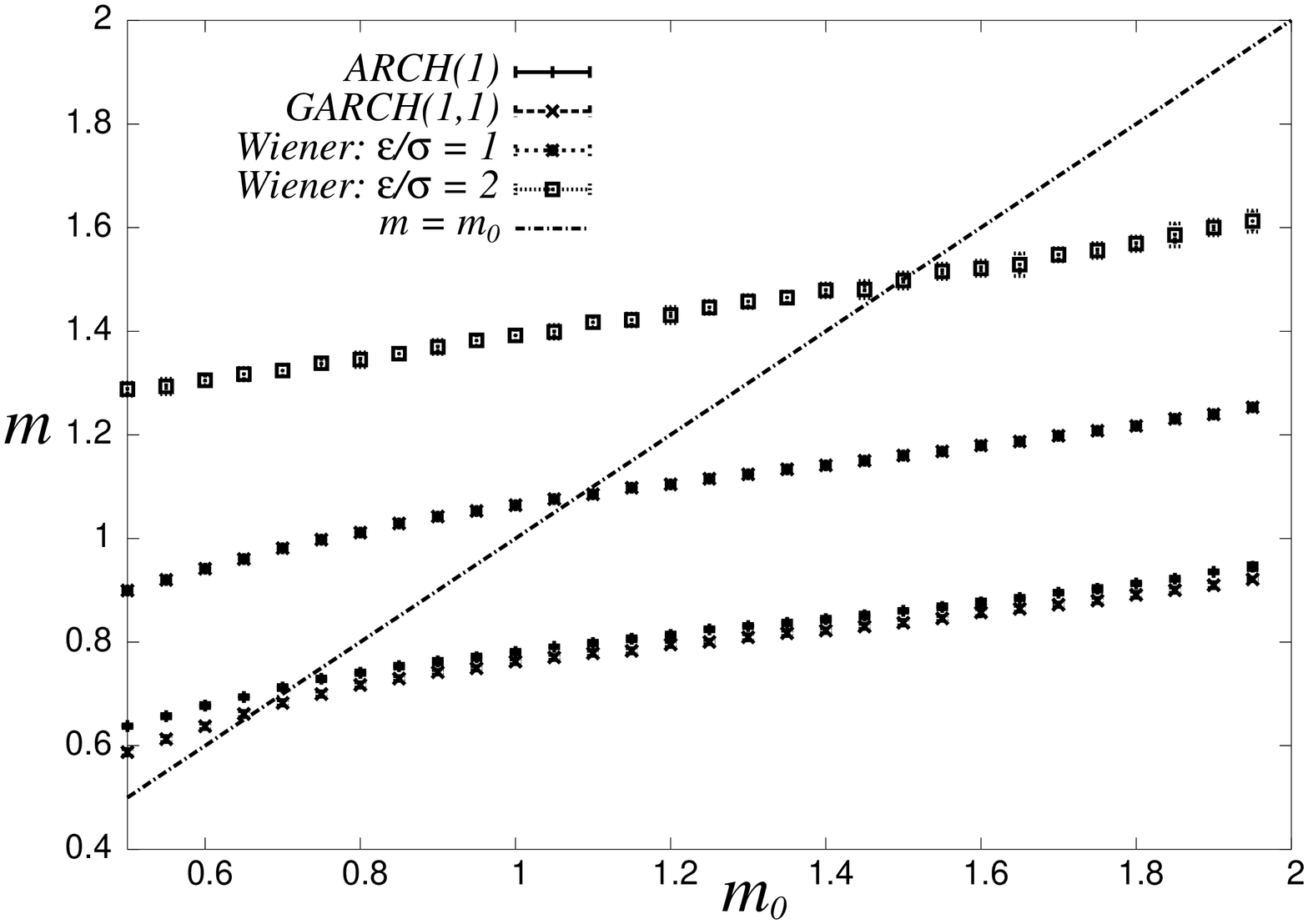}}
\rotatebox{-90}{\includegraphics[width=5.6cm]{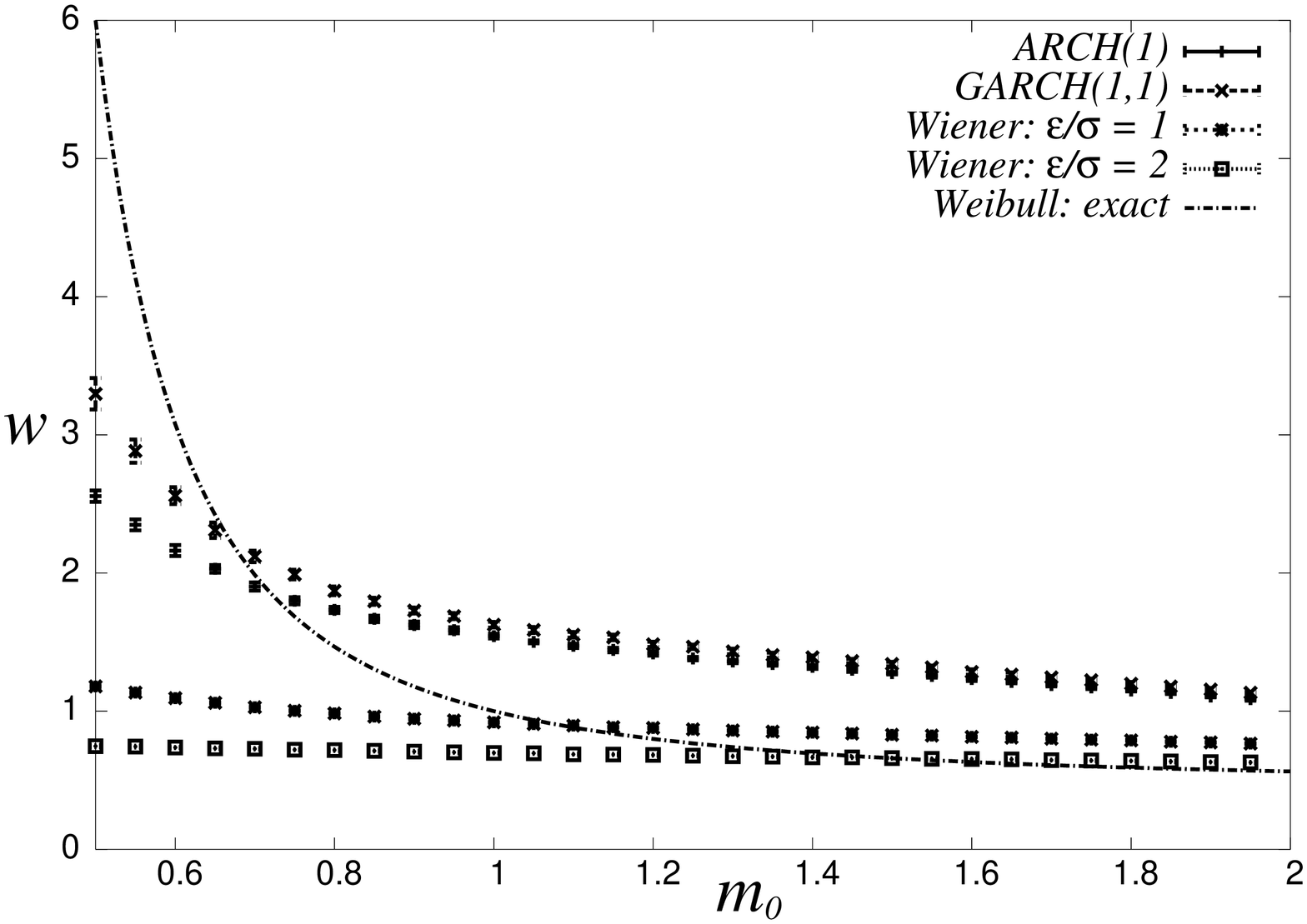}}
\end{center}
\caption{\footnotesize 
The $m_{0}$-$m$ relations 
for the Wiener process and 
the ARCH(1) and GARCH(1,1) models (left). 
The right panel 
shows the $m_{0}$-dependence of the 
average waiting time $w$ for 
each 
stochastic process. 
In this panel, 
the unit of the vertical axis is second. 
Each error-bar is evaluated by 
$20$ independent runs.}
\label{fig:fg3}
\end{figure}
In the left panel of FIG. \ref{fig:fg3}, 
we plot the 
$m_{0}$-$m$ relations 
for the stochastic processes: 
the Wiener process and 
the ARCH(1) and GARCH(1,1) models. 
In this panel, 
the crossing point between the 
line $m=m_{0}$ and 
each plot 
indicates 
the value of $m_{0}$ for which 
the distribution $P(t)$ of 
the first-passage time 
remains the same as that of 
the time interval $\Delta t$ 
of the raw data $P_{W}^{0}(\Delta t)$.  
Therefore, 
below the line $m=m_{0}$, 
the rate window 
affects the 
stochastic process 
so as to 
decrease the parameter $m_{0}$ of 
the Weibull distribution, 
whereas, 
above this line, 
the parameter $m_{0}$ increases 
to $m \,(>m_{0})$. 
As mentioned earlier, 
empirical data analysis 
of the Sony Bank rate \cite{Sazuka2} 
suggested that 
the waiting time or 
the time interval 
$\Delta t$ of the 
Sony Bank rate obeys the Weibull 
distribution with $m=0.59$. 
Therefore, 
there is a possibility 
that the raw data 
before applying the rate window 
could be modelled by 
the GARCH(1,1) model. 
The right panel of FIG. \ref{fig:fg3} 
shows the $m_{0}$-dependence of 
the average waiting time 
obtained by 
the data 
after the rate 
window filter is applied. 
From this panel, 
we find that 
the GARCH(1,1) model 
reproduces the average waiting time $w$ for 
the raw data (Weibull: exact) 
more effectively than the other three 
models. 
\section{Evaluation of expected rewards}
\begin{figure}[ht]
\begin{center}
\rotatebox{-90}{\includegraphics[width=8cm]{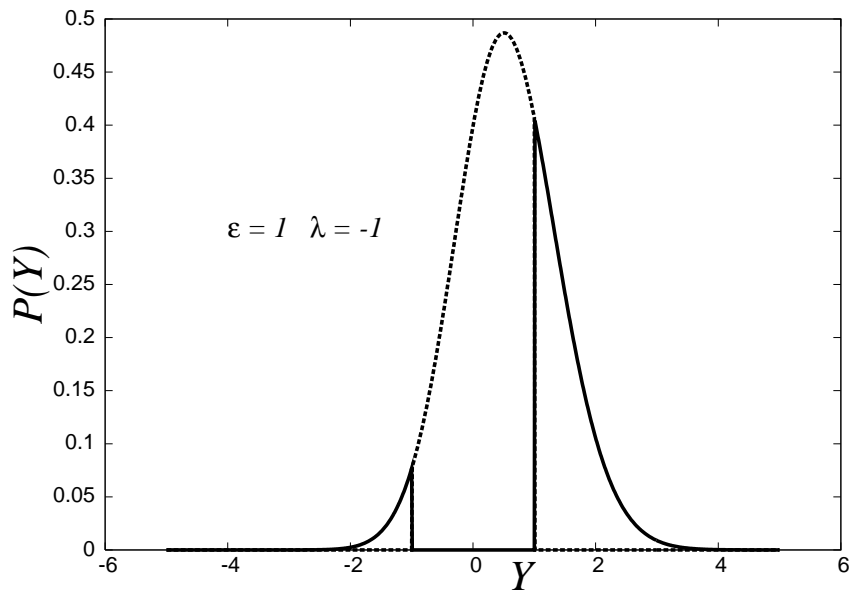}}
\end{center}
\caption{\footnotesize 
The skew-normal distribution 
defined by 
(\ref{eq:skew-normal}) 
with $\epsilon=1$ and 
$\lambda=-1$. 
The unit of the horizontal axis is yen.}
\label{fig:fg7}
\end{figure}

In the previous sections, 
we considered 
the average waiting time 
that a customer has 
to wait until the next update of the Sony Bank rate after 
she or he logs in to her/his computer systems. 
To evaluate 
the average waiting time, we used 
the renewal-reward theorem, 
which is wellknown 
in the field of queueing theory \cite{Tijms,Oishi}. 
Besides the average 
waiting time, 
we need to consider another relevant quantity 
while investigating 
the statistical properties of the Sony Bank rate 
from a different viewpoint. 
For instance, 
the cumulative return that the 
customers can 
be expected to obtain  
during the time interval $t$, 
\begin{eqnarray}
R(t) & = & 
\sum_{n=1}^{N(t)}
Y_{n},\,\,\,
Y_{n} = 
X_{n+1}-X_{n},
\label{eq:reward}
\end{eqnarray}
is one of the candidates for 
such relevant quantities. 
In the 
definition 
of the cumulative return  
(\ref{eq:reward}), 
$N(t)$ refers to the number of 
rate changes within time 
interval $t$. 
The return 
$Y_{n}$ is the difference between 
the rates of 
two successive time stamps  
$n+1$ and $n$. 
Then, 
the long-time average 
of the cumulative return 
defined 
by $R \equiv \lim_{t \to \infty}
(R(t)/t)$ is 
rewritten as 
\begin{eqnarray}
R & = & 
\lim_{t \to \infty}
\frac{R(t)}{t}
= 
\lim_{t \to \infty}
\frac{N(t)}{t}
\cdot 
\frac{R(t)}{N(t)}.
\end{eqnarray}
Taking into account 
the following 
relation : 
\begin{eqnarray}
\lim_{t \to \infty}
\frac{N(t)}{t} & = &  
E(t)^{-1}
\end{eqnarray}
and the law of 
large numbers 
$R(t)/N(t)={\cal E}(Y)$, 
we obtain 
a long-time average of the cumulative 
return, i.e. the {\it reward rate} 
$R$ as follows: 
\begin{eqnarray}
R & = & 
\frac{{\cal E}(Y)}
{E(t)},
\end{eqnarray}
where 
we define 
$E(t)$ as the 
expectation of 
the first-passage time 
$t$ and 
${\cal E}(Y)$ as 
the average of the return 
$Y$ over the 
probability distribution $P(Y)$. 
If we set 
${\cal E}(Y)=\mu$ and 
assume that the first-passage 
time $t$ obeys the Weibull distribution, 
the above reward rate 
$R$ is given by 
\begin{eqnarray}
R & = & 
\frac{\mu m}
{a^{1/m}
\Gamma 
\left(
\frac{1}{m}
\right)}.
\end{eqnarray}
Obviously, 
if the distribution of 
the difference of the rates 
$Y$ between two arbitrary successive time stamps 
obeys symmetric 
distribution around 
$Y=0$, 
$\mu$ becomes zero, 
and as a result, 
the reward rate $R$ also becomes zero. 
For performing 
theoretical evaluations for $R$, namely, 
to investigate the skewness $S$-dependence of 
the reward rate $R$, 
we assume that 
the stochastic variable $Y$ 
obeys the following 
skew-normal distribution, 
wherein 
the regions 
$(-\infty, -\epsilon], 
[\epsilon, \infty)$ 
are cut off: 
\begin{eqnarray}
P(Y) & = & 
\sqrt{
\frac{2}{\pi}}\,
{\rm e}^{-\frac{Y^{2}}{2}}
H(\lambda Y) 
\Theta (|Y|-\epsilon), 
\label{eq:skew-normal}
\end{eqnarray}
where 
$\Theta(x)$ is a 
Heaviside 
step function and 
$H(x)$ is 
defined by 
$H(x)=\int_{x}^{\infty}
(dt/\sqrt{2\pi})\, 
{\rm e}^{-t^{2}/2}$. 
The variable $Y$ cannot take any values 
lying in the 
intervals 
$(-\infty,-\epsilon]$ and $[\epsilon, \infty)$ 
because of 
the method of generating of the Sony Bank rate, as 
we have already explained in the previous sections. 
That is, 
the Sony Bank rate 
changes if and only 
if the 
difference $|Y_{n}|=
|X_{n+1}-X_{n}|$ 
becomes larger than 
the width of the 
rate window $\epsilon$. 
In FIG. \ref{fig:fg7}, 
we show the 
skew-normal 
distribution 
(\ref{eq:skew-normal}) for 
$\epsilon=-1$ and 
$\lambda=1$. 
Note that $P(Y)$ becomes 
a normal distribution for the limits $\lambda \to 0$ and 
$\epsilon \to 0$. 

For this skew-normal distribution 
(\ref{eq:skew-normal}), 
we easily obtain the average $\mu = 
\int_{-\infty}^{\infty}YP(Y)$ as 
\begin{eqnarray}
\mu & = & 
\sqrt{
\frac{2}{\pi}
}\,
{\rm e}^{-\frac{\epsilon^{2}}{2}}
\left\{
2H(\lambda \epsilon) -1
\right\}
- 
2\sqrt{
\frac{2}{\pi}}
\frac{\lambda}
{\sqrt{1+\lambda^{2}}}
H(\epsilon
\sqrt{1+\lambda^{2}}). 
\end{eqnarray}
The second 
and 
the third moments 
$\mu_{2}=
\int_{-\infty}^{\infty}
Y^{2}P(Y), 
\mu_{3}=
\int_{-\infty}^{\infty}
Y^{3}P(Y)$ 
lead to 
\begin{eqnarray}
\mu_{2} & = & 
\frac{\epsilon}
{\sqrt{2\pi}}\,
{\rm e}^{-\frac{\epsilon^{2}}{2}}
+ 
H(\epsilon), \\
\mu_{3} & = & 
2\epsilon^{2} 
\sqrt{
\frac{2}{\pi}}\,
{\rm e}^{-\frac{\epsilon^{2}}{2}}
H(\lambda \epsilon) + 
4
\sqrt{\frac{2}{\pi}}\,
{\rm e}^{-\frac{\epsilon^{2}}{2}}
H(\lambda \epsilon) - 
4
\sqrt{\frac{2}{\pi}}
\frac{\lambda}{\sqrt{1+\lambda^{2}}}
H(\epsilon 
\sqrt{1+\lambda^{2}}) \nonumber \\
\mbox{} & - & 
2\sqrt{\frac{2}{\pi}}
\lambda
(1+\lambda^{2})^{-3/2}
\left\{
\sqrt{
\frac{1+\lambda^{2}}{2\pi}
}
\, 
\epsilon \, 
{\rm e}^{-\frac{\epsilon^{2}}{2}}
+ 
H(\epsilon 
\sqrt{1+\lambda^{2}}) 
\right\}
- 
\sqrt{
\frac{2}{\pi}}\,
{\rm e}^{-\frac{\epsilon^{2}}{2}}
(2+\epsilon^{2}).
\end{eqnarray}
The skewness $S$ of 
the distribution 
$P(Y)$ is 
written in terms of 
these moments as follows: 
\begin{eqnarray}
S & = & 
\frac{{\cal E}(
(Y-\mu)^{3})}
{\sigma^{3}} = 
\frac{\mu_{3}-
3\mu \mu_{2} + 
2\mu^{3}}
{\sigma^{3}}, 
\end{eqnarray}
where 
$\sigma$ is the standard deviation 
$\sigma 
\equiv 
\sqrt{\mu_{2}^{2} -\mu^{2}}$. 

\begin{figure}[ht]
\begin{center}
\rotatebox{-90}{\includegraphics[width=8cm]{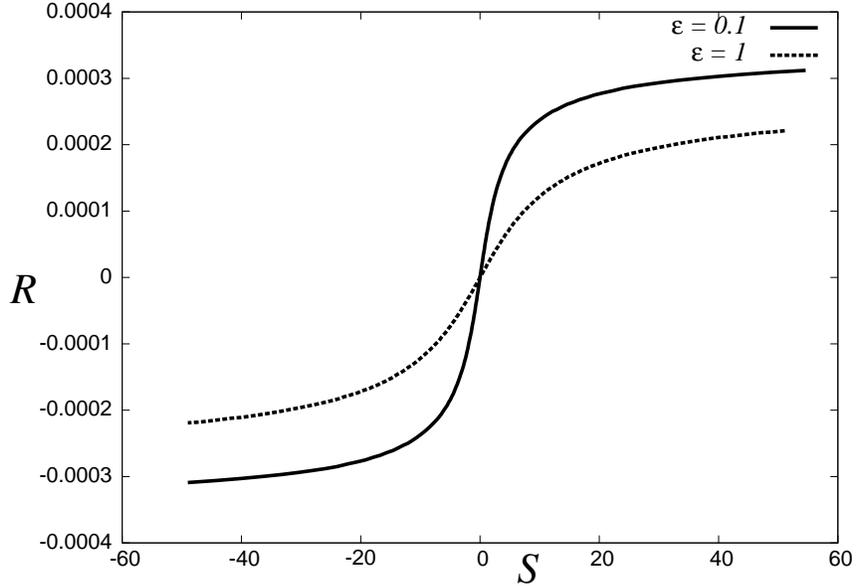}}
\end{center}
\caption{\footnotesize 
The reward rate $R$ 
as a function of 
the skewness $S$ of the skew-normal 
distribution. 
The unit of the vertical axis is 
yen per second. 
We set the parameters $m$ and $a$ for 
the Weibull distribution 
with respect to 
the first-passage time 
as $(m,a)=(0.59,50.855)$. 
We plot it for the cases of 
$\epsilon=0.1$ and $1$ with $\lambda=-1$. }
\label{fig:fg8}
\end{figure}

In following, 
we evaluate the reward 
rate $R$ 
as a function of 
the skewness $S$ for the 
parameter values 
$m=0.59$ and $a=50.855$. 
In FIG. \ref{fig:fg8}, 
we plot them 
for the cases of 
$\epsilon=0.1$ and $1$. 
From this figure, we 
find that 
the reward rate increases 
dramatically as the skewness of 
the skew-normal distribution 
(\ref{eq:skew-normal}) becomes positive. 
As the skewness $S$ increases, 
the reward rate $R$ 
saturates at 
$0.0003$ [yen/s]
for our model system, 
wherein the time interval of the rate change 
$t$ obeys 
the Weibull distribution 
with $(m,a)=(0.59,50.855)$, 
and the difference of 
the rate $Y$ 
follows 
the skew-normal distribution 
(\ref{eq:skew-normal}). 
In this figure, 
we also find that 
if we increase the width $\epsilon$ 
of the rate window from 
$\epsilon=0.1$ to $1$, 
the reward rate $R$ 
decreases. 
For companies or 
internet banks, 
this kind of information 
might be useful because 
they can control the reward rate (this takes both 
positive and negative values) 
for their customers by 
tuning the width $\epsilon$ of the 
rate window in their computer simulations.

Moreover, 
we should note that we can also 
evaluate the expected reward 
$\overline{R}=Rw$, which is the return 
that the customers can expect 
to encounter after they 
log in to their computer systems. 
By combining 
the result obtained 
in this section 
$R=0.0003$ [yen/s] 
and the average waiting time 
$w=42.236 \times 60=2534.146$ [s], 
we 
conclude that 
the expected reward should be 
smaller than 
$\overline{R}=Rw \,\sim\,0.76$ [yen]. 
This result seems to be important 
and useful for both the customers and 
the bank, e.g. in setting the transaction cost. 
Of course, 
the probabilistic model considered here
for $P(t)$ or 
$P(Y)$ is just an example for  
explaining the stochastic 
process of the real or empirical rate change. 
Therefore, 
several modifications 
are needed to conduct a  
much deeper investigation 
from theoretical viewpoint. 
Nevertheless, 
our formulation might 
be particularly useful in dealing with 
price changes in real financial markets. 

\section{Summary and discussion}

In this paper, we introduced the concept of 
queueing theory to 
analyse price changes in a financial market, 
for which, 
we focus on the USD/JPY exchange rate of 
Sony Bank, 
which is an internet-based bank. 
Using the renewal-reward theorem and 
on the assumption that the Sony Bank rate 
is described by a first-passage process 
whose FPT distribution follows a Weibull distribution, 
we evaluated the average waiting time 
that Sony Bank customers have to wait until 
the next rate change after they 
log in to their computer systems. 
The theoretical prediction and 
the result from the empirical data analysis 
are in good agreement on the value of the 
average waiting time. 
Moreover, our analysis revealed that 
if we assume that the Sony Bank rate is 
described by a Poisson arrival process with 
an exponential FPT distribution, 
the average waiting time predicted by the renewal-reward 
theorem is half the result predicted by the empirical 
data analysis.  
This result justifies the non-exponential time 
intervals of the Sony Bank USD/JPY exchange rate. 
We also evaluated the reward 
that a customer could be expected to encounter 
by the next price change 
after they log in to their computer systems. 
We assumed that the return and FPT follow 
skew-normal and Weibull distributions, respectively,  
and found that the expected return for 
Sony Bank customers is smaller than $\sim\, 0.76$ yen. 
This kind of information 
about statistical properties might be useful for 
both the costumers and 
bank's system engineers. 

As mentioned earlier, in this paper, 
we applied queueing theoretical analysis to the Sony Bank 
rate, which is generated as a first-passage process with 
a rate window of width $2\epsilon$. 
We did not mention the high-frequency 
raw data underlying behind the Sony Bank rate 
because Sony Bank does not 
record raw data and the data itself is not available to us. 
Although our results did not suffer from this lack of 
information, 
the raw data seems to be attractive and interesting 
as a material for financial analysis. 
In particular, the effect of the rate window on 
high-frequency raw data should be investigated empirically. 
However, it is impossible for us to conduct such an 
investigation for the Sony Bank rate. 
To compensate for this lack of information, 
we performed the GARCH simulations, wherein 
the duration between price changes 
in the raw data obeys a Weibull distribution with 
parameter $m_{0}$. 
Next, we investigated the effect of the rate window 
through the first-passage time distribution 
under the assumption that it 
might follow a Weibull distribution 
with parameter $m \, (\neq m_{0})$. 
The $m_{0}$-$m$ plot was thus obtained. 
Nevertheless, an empirical data analysis to investigate 
the effect might be important. 
Such a study is beyond the scope of this paper; 
however, it is possible for us to generate the 
first-passage process from other high-frequency raw data 
such as BTP futures. 
This analysis 
is currently under way and 
the results will be reported in our forthcoming article
\cite{ISE}. 

As mentioned in TABLE I, the amount of data 
per day for the Sony Bank rate is about $70$ points,  which is 
less than the tick-by-tick high-frequency data. 
This is because the Sony Bank rate is generated as 
a first-passage process of the raw data. This means 
that the number of data is too few to confirm whether 
time-varying behaviour is actually observed. 
In our investigation of the limited data, 
we found that the first-passage time distribution in a specific 
time regime (e.g. Monday of each week) obeys a Weibull distribution, 
but the parameter $m$ is slightly different from $0.59$. 
However, this result has not yet confirmed because of 
the low number of data points. Therefore, 
we used the entire data (about $31,000$ 
points from September 2002 to May 2004) 
to determine the Weibull distribution. 
In this paper, we focused on the evaluation 
of the average waiting time and achieved  
the level of accuracy mentioned above. 
In addition, if Sony Bank records more data in future, 
we might resolve this issue, namely, whether 
the Sony Bank rate exhibits time-varying behaviour. 
This is an important area for future 
investigation.

Although we dealt with the Sony Bank USD/JPY exchange rate 
in this paper, our approach is general and 
applicable to other 
stochastic processes 
in financial markets. 
We hope that it is widely used to 
evaluate various useful statistics in 
real markets.
\begin{acknowledgments}
We thank Enrico Scalas for fruitful discussion and 
useful comments. 
J.I. was financially supported 
by the {\it Grant-in-Aid for Young Scientists (B) 
of The Ministry of Education, Culture, 
Sports, Science and Technology (MEXT)} 
No. 15740229. and the {\it Grant-in-Aid 
Scientific Research on Priority Areas 
``Deepening and Expansion of Statistical Mechanical Informatics (DEX-SMI)" 
of The Ministry of Education, Culture, 
Sports, Science and Technology (MEXT)} 
No. 18079001. 
N.S. would like to acknowledge useful discussion 
with Shigeru Ishi, President of Sony Bank.

\end{acknowledgments}

\end{document}